\documentclass[final,1p,number,sort&compress]{elsarticle}

\usepackage{amssymb}
\usepackage[utf8]{inputenc}
\usepackage{amsmath}
\usepackage{lipsum}
\usepackage[labelfont=bf]{caption}
\usepackage{subcaption}
\usepackage{multirow}
\usepackage{mathtools}
\usepackage{xfrac}
\usepackage[referable]{threeparttablex}
\usepackage{footnote}
\usepackage{pdflscape}
\usepackage[colorlinks]{hyperref}
\usepackage{makecell}

\usepackage{tikz}
\newcommand*\circled[1]{\tikz[baseline=(char.base)]{
            \node[shape=circle,draw,inner sep=1pt] (char) {#1};}}

\journal{Nuclear Instruments and Methods in Physics Research}

\begin{document}

\begin{frontmatter}

\title{Phase-Imaging Ion-Cyclotron-Resonance Mass Spectrometry with the Canadian Penning Trap at CARIBU}

\author[UMB,ANL]{D. Ray \corref{cor1}} 
\author[UMB,ANL]{A.A. Valverde}
\author[NDU]{M. Brodeur}
\author[McGill]{F. Buchinger}
\author[ANL,UMB]{J.A. Clark}
\author[NDU,ANL]{B. Liu}
\author[LSU,ANL]{G.E. Morgan}
\author[LBNL]{R. Orford}
\author[NDU]{W.S. Porter}
\author[ANL,UC]{G. Savard}
\author[UMB]{K.S. Sharma}
\author[ANL,IMPC]{X.L. Yan}

\address[UMB]{Department of Physics and Astronomy, University of Manitoba, Winnipeg,  MB, R3T 2N2, Canada}

\address[ANL]{Physics Division, Argonne National Laboratory, Lemont, IL 60439, USA}

\address[NDU]{Department of Physics and Astronomy, University of Notre Dame, Notre Dame, IN 46556, USA}

\address[McGill]{Department of Physics, McGill University, Montreal, QC H3A 2T8, Canada}

\address[LSU]{Department of Physics and Astronomy, Louisiana State University, Baton Rouge, LA 70803, USA}

\address[LBNL]{Nuclear Science Division, Lawrence Berkeley National Laboratory, Berkleley, CA 94720, USA}

\address[UC]{Department of Physics, University of Chicago, Chicago, IL 60637, USA}

\address[IMPC]{Institute of Modern Physics, Chinese Academy of Sciences, Lanzhou 730000, China}

\cortext[cor1]{Corresponding author: dray@triumf.ca; Current Address: TRIUMF, 4004 Wesbrook Mall, Vancouver, BC V6T 2A3, Canada}

\begin{abstract}
The Canadian Penning Trap mass spectrometer (CPT) has conducted precision mass measurements of neutron-rich nuclides from the CAlifornium Rare Isotope Breeder Upgrade (CARIBU) of the Argonne Tandem Linac Accelerator System (ATLAS) facility at Argonne National Laboratory using the Phase-Imaging Ion-Cyclotron-Resonance (PI-ICR) technique for over half a decade. Here we discuss the CPT system,  and methods to improve accuracy and precision in mass measurement using PI-ICR including some optimization techniques and recently studied systematic effects.
\end{abstract}


\end{frontmatter}

\section{Introduction}\label{sec:intro}

Penning traps have been the most precise instruments for mass measurements of stable and radioactive nuclides for many different fields of physics for over four decades~\cite{gartner_first_penning_mass_1978, KLUGE_penning_mass_spec_201326, blaum_mass_spec_2006}.  Penning trap mass spectrometry relies on determining the cyclotron frequency ($\nu_{c}$) of an ion of mass $M$ and charge state $q$ in a magnetic field $B$:
\begin{equation}\label{eq:penning}
    \nu_{c} = \frac{q e}{2 \pi M} B \; ,
\end{equation}
where $e$ is the charge of an electron. Once the $\nu_{c}$ of a target ion species and a calibrant (Cal) ion species of known mass (for calibrating the magnetic field) are precisely measured, the atomic mass ($m$) of the target species is determined as, 
\begin{equation}\label{eq:mass}
    m = \frac{q}{q^{\text{Cal}}} \; r \; (m^{\text{Cal}} - q^{\text{Cal}} m_{e} + B_{e}^{\text{Cal}}) + q m_{e} - B_{e}\; ,
\end{equation}
where $r=\sfrac{\nu_{c}^{\text{Cal}}}{\nu_{c}}$ is the $\nu_{c}$ ratio between the calibrant and the target ion species, $m_e$ is the mass of an electron, and $B_{e}^{\text{Cal}}$ and $B_{e}$ are the electron binding energies of the calibrant and the target species. For singly or doubly charged ions, $B_{e}^{\text{Cal}}$ and $B_{e}$ are of the order of 1~-~10s of eV, over an order of magnitude smaller than the usual precision achieved in the technique described here, and can therefore be ignored during mass determination.
 
The Canadian Penning Trap (CPT)~\cite{guy_cpt_first_1997, guy_cpt_anl_first_2001} is located at Argonne National Laboratory (ANL) where it has conducted precision mass measurements of neutron-rich nuclides from the CAlifornium Rare Isotope Breeder Upgrade (CARIBU)~\cite{guy_caribu_short_1_2008} at the Argonne Tandem Linac Accelerator System (ATLAS) facility, using the Phase Imaging Ion-Cyclotron-Resonance (PI-ICR)~\cite{eliseev_piicr_APB_long_2014, orford_nimb_piicr_cpt_2020, Welker:2641361, piicr_jyfltrap_2021, piicr_trigatrap_2023, FRIB_piicr_PhysRevLett.132.152501} technique. 
This paper discusses the PI-ICR operations, and some new systematic studies which have enabled the CPT to achieve ppb-level precision in mass measurements.

\section{The CARIBU facility}\label{subsec:caribu}

CARIBU is a unique facility which produces neutron-rich nuclides by collecting spontaneous fission fragments from a $^{252}$Cf source,  and sends them as low-energy beams either to the CARIBU low-energy experimental area where the CPT resided or the newly constructed Area 1, or to the Electron Beam Ion Source (EBIS)~\cite{vondrasek2012ebis} for charge-breeding and acceleration before injection into ATLAS for other experiments. At the heart of CARIBU is a $\sim$0.5~Ci $^{252}$Cf source decaying with a half-life of $2.645(8)$~years and having a spontaneous fission branch of 3.1028(7)\%~\cite{nubase2020_Kondev2021}. The fission fragments pass through a gold degrader foil and enter a large volume gas catcher~\cite{guy_caribu_gas_catcher_short_1_2016}
consisting of a cylindrical body and a cone. Here they are stopped and thermalized in high-purity He gas, and are extracted using a combination of gas flow, radio-frequency (RF) fields and direct current (DC) fields as singly or doubly charged ions. The gas catcher and associated radio-frequency quadrupole (RFQ) ion guides sit on a $36$~kV high voltage (HV) platform. The extracted ions first undergo mass-selection on the basis of their mass numbers ($A$) and $q$ by passing through the isobar separator~\cite{caribu_isobar_separator_1_2008} consisting of a pair of dipole magnets and operated at a resolving power of 5,000, before they are cooled and bunched using a RFQ cooler-buncher. The bunched beam is extracted at $3$~keV every 50~ms or 100~ms, and then gets further mass-resolved by a multi-reflection time-of-flight (MR-TOF) mass separator~\cite{hirsh_caribu_mrtof}. For most CPT experiments, the ions are cycled inside the MR-TOF for 15~-~20~ms, enabling achievement of a mass resolution of 100,000.
At the exit of the MR-TOF, the mass-resolved beam passes through a Bradbury-Nielsen gate (BNG), where the target nuclides are precisely selected on the basis of their TOF and sent to the CPT system. 

All the beamline components downstream from the buncher require triggering at the precise times when the bunched beam approaches them. The trigger signals are all initiated by the ions' ejection from the buncher. The signals for all the components except the BNG are fine-tuned to the $A/q$ of the ions in the beam, while the BNG trigger is fine-tuned to the precise $m/q$ of the target nuclide.

\section{The CPT system}\label{sec:cpt_sys}

\begin{figure}
\centering
\includegraphics[clip,width=0.99\textwidth]{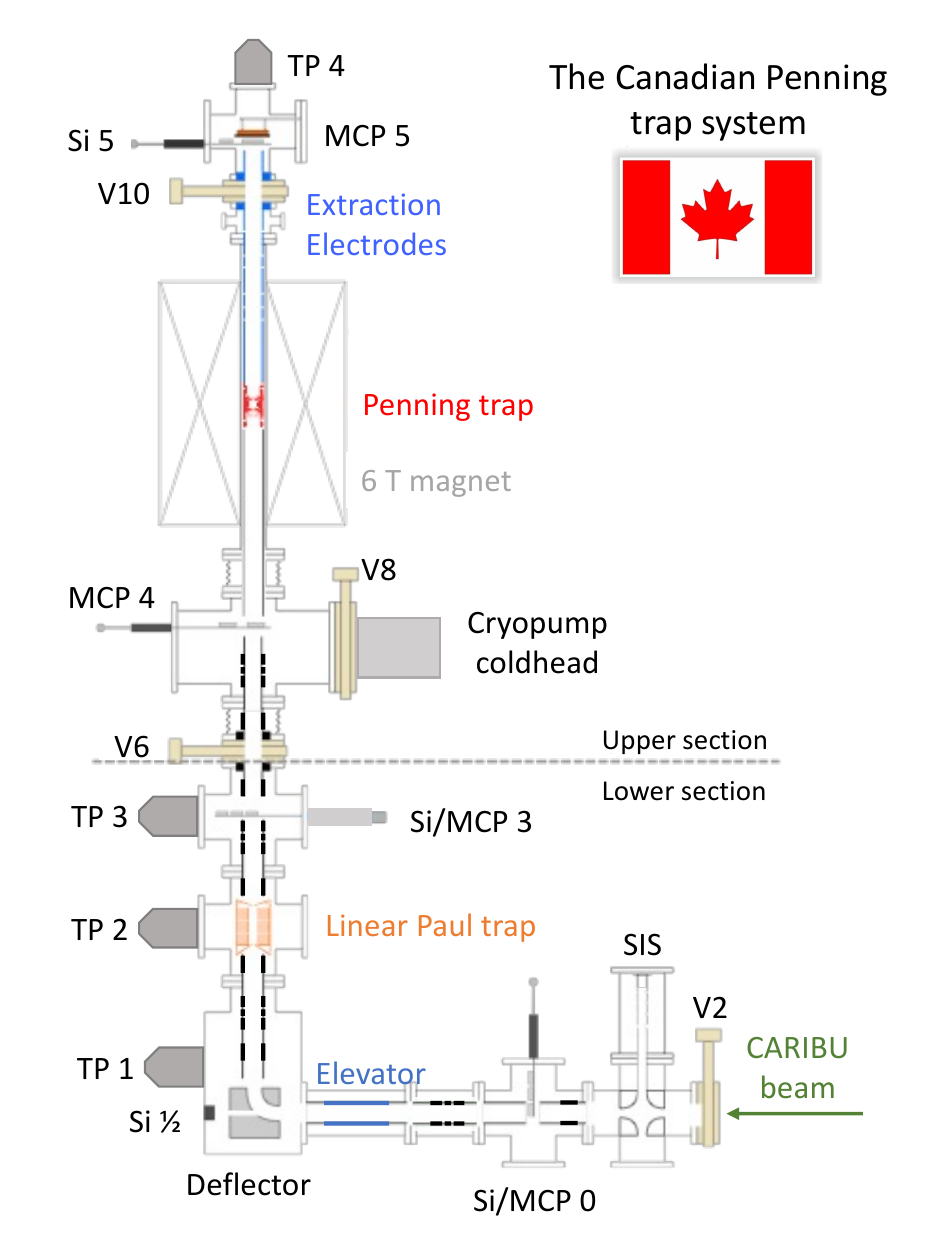} 
\captionsetup{width = 12.5cm}
\caption[The CPT Tower]{\label{fig:cpt_tower}Schematic diagram of the 4.5 m CPT tower (not to scale), showing the major components.}
\end{figure}

\subsection{Overview}\label{subsec:overview}

The CPT system consists mainly of two sections, a stable-ion-source (SIS), an elevator drift tube and a linear Paul trap (LT) constituting the lower, beam-preparation section, and a precision Penning trap, the CPT, equipped with a position-sensitive microchannel plate detector (PS-MCP, also known as MCP 5) making up the upper measurement section.  The sections are housed in a $\sim$4.5~m tower structure as shown in Fig.~\ref{fig:cpt_tower}.  The system also includes a host of other components like diagnostic silicon surface barrier (Si) and MCP detectors, turbomolecular pumps (TP) and standard electrostatic beamline components,  some of which have been excluded from the figure for simplicity.  The lower section is maintained at high vacuum (10~$\mu$Torr to 100~nTorr) while the upper section is maintained at ultra-high vacuum ($< 1$~nTorr) as required for precision measurement. The whole system can be isolated from CARIBU using a gate valve V2, while V6 separates the upper and lower sections.

\subsection{Offline stable ion source (SIS)}\label{subsec:offline_source}

The SIS is located at the bottom of the tower between the valve V2 and the first diagnostic station of the CPT system ``Si/MCP 0'', as can be seen in Fig.~\ref{fig:cpt_tower}. It consists of a mixed-alkaline, surface-ionization ion source (Model 101139) from Heat Wave Labs~\cite{SIS_details_page},  containing salts of natural-abundance Cs, Rb and K in roughly equal proportions, and heated using a pair of coaxial heating elements. This produces continuous ion streams of $^{133}$Cs$^{+}$, $^{85,87}$Rb$^{+}$, and $^{39,41}$K$^{+}$, which are accelerated into the main beamline, and used for optimizing the different beamline components and for magnetic-field calibration during precision measurement campaigns due to their precisely known masses~\cite{ame2020_b_Wang2021}.

\subsection{Beam preparation : Elevator and linear Paul trap}\label{subsec:cpt_prep}

The traps (LT and CPT) are designed to operate at low voltages ($\sim \pm$20~V), and to trap and manipulate ions with energies of a few eVs.
Therefore, the 3~keV energy of the incoming beam from CARIBU needs to be lowered.  This is accomplished using an elevator that consists of a 30~cm long drift tube. A signal mirroring the BNG trigger is used to trigger a delayed pulse which lowers the potential of the elevator from ground to ${-}$3~kV when the ion bunch is inside it.  Ions from the SIS are extracted at the transport voltage of the tower and do not require the elevator. Therefore, the elevator voltage is maintained at the standard transport voltage of the system for SIS beam.

The LT is a gas-filled linear RFQ Paul trap, used as a buffer trap to make operation of the system more general. The segmented RFQ electrode structure is cooled by circulating liquid nitrogen, using a custom-made pump~\cite{LN2_pump_201635_Li_Caldwell}, to reduce thermal excitations, and it is filled with high-purity He gas at 10~$\mu$Torr. After injection into the LT, ions are trapped, using a combination of radial RF fields and an axial DC potential well, and cooled via collisions with the He gas. The RF excitations are applied from a Stanford Research System (SRS), model DS345 function generator, through an amplifier.
Ion bunches are stored in the LT for a minimum of 30~ms reducing their emittance, before they are ejected towards the CPT.

\subsection{Penning trap}\label{subsec:cpt}

\begin{figure}
\centering
\includegraphics[clip,width=0.99\textwidth]{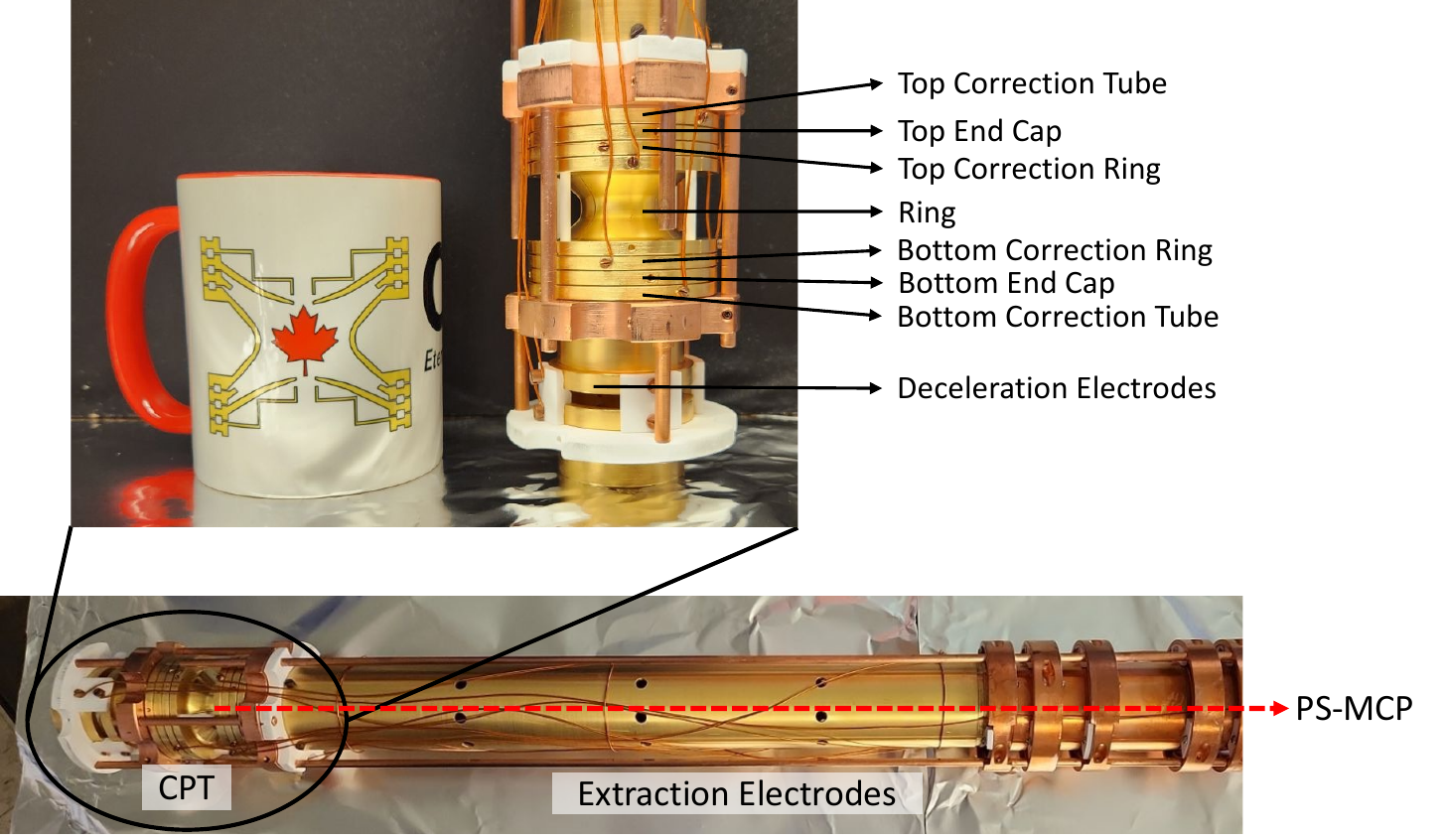}
\captionsetup{width = 12.5cm}
\caption[The CPT assembly]{\label{fig:cpt}The CPT electrode assembly showing the trap (zoomed in on the inset, with a mug added for scale) and the extraction electrodes. An ion's path from the trap towards the PS-MCP is shown with the red dotted line.}
\end{figure}

The Canadian Penning Trap~\cite{guy_cpt_first_1997} is a hyperboloid Penning trap with characteristic dimensions of $r_{0}=11.6$ mm and $z_{0} = 10$ mm consisting of ten electrodes: pairs (top and bottom) of correction tubes, end caps and correction rings, on either side of a 4-segment ring electrode. There are two 5~mm diameter apertures on both sides through the end cap and correction tube electrodes for entry and exit of the ions. The correction electrodes are used to compensate for the non-ideal or practical aspects in the trap design, like the segmented ring, the finite size of the electrodes and the apertures, and create a harmonic electric potential near the center of the trap.
Typical DC biases applied to the seven CPT electrodes during capture, trapping and ejection of ions are shown in Tab.~\ref{tab:cpt_dc_biases}.
The trap sits inside the bore of a 6~T superconducting magnet and forms one physical assembly together with a pair of deceleration injection electrodes upstream and nine cylindrical extraction drift tubes downstream (called TOF A to I)~\cite{guy_cpt_anl_first_2001}. This is shown in Fig.~\ref{fig:cpt}. All the trap electrodes are made with gold-plated, oxygen-free high-purity copper (OFHC) with Macor$^{\text{\circled{R}}}$ insulators to satisfy the electric conductivity, magnetic susceptibility and UHV requirements. 

During the transfer between the LT and the CPT, some energy spread is imparted in the ion cloud. Therefore once the ions are captured, an evaporative cleaning procedure is conducted to remove the high energy ions, minimize the axial spread, and ensure that the trapped cloud sits at the center of the trap, always probing the same magnetic field. Next, RF excitations are applied to the ring electrode as required for mass measurement using PI-ICR, after which the ions are ejected from the trap.  The capture, ejection and evaporative cleaning are done using a series of SRS DS345 function generators, controlled by a CAMAC-based control system. The function generators are programmed to output user-defined waveforms, which are combined to the trapping voltages, and applied to the electrodes.
A dedicated four channel arbitrary waveform generator from Tabor Electronics (model WW1074)~\cite{tabor_1074} is used to provide the PI-ICR RF excitations to the four segments of the ring electrode. 
An external 10 MHz clock ensures that all four
channels are synced together.

PI-ICR depends on accurate projections of the ions' in-trap positions on to the PS-MCP. To minimize distortions and optimize the quality of this projection, the ions’ TOF through the magnetic field gradient from the trap to the detector had to be minimized. As part of the ejection procedure, the top end cap and correction tube electrodes are pulsed to eject the ions from the trap, and the TOF A electrode, which is the extraction electrode situated closest to the trap and in a region of large magnetic field gradient, is pulsed from $-$20~V to $-$300~V thereby accelerating the ions without perturbing the harmonicity of the trap. The rest of the drift tubes (TOF B~-~I except F) are maintained at a fixed negative potential of $-$400~V as the ions are transferred towards the PS-MCP. TOF F is maintained at $-$850~V to focus the ions on the PS-MCP. 
The biases to the electrodes TOF B~-~I are tuned to optimize the image on the PS-MCP and minimize distortions. The exact value of the magnification factor could not be determined because of a non-circular projection from the trap to the detector, as discussed in Sec.~\ref{subsubsec:elliptical projection}. All measurements are conducted at a radius of $\sim$5 mm, which is the maximum radius of the ions after the $\nu_{+}$ excitation (that is part of the PI-ICR measurement technique discussed in Sec.~\ref{subsec:piicr cpt})
 observed on the detector.
The detector setup is a commercial package from RoentDek Handels GmbH~\cite{PS_MCP_manual}, and includes the PS-MCP (DLD40), consisting of a pair of image quality MCP plates (front and back),  with a delay line anode, along with associated electronics comprising of a signal decoupling feedthrough (FT12TP), a fast signal amplifier (FAMP6), a constant fraction discriminator (CFD4c), a high resolution time-to-digital converter (TDC8HP), and read-out software (CoboldPC).  Each MCP plate has an active diameter of 45~mm and thickness of 1.5~mm. The set up provides an electron gain of $10^{7}$ for a voltage difference of $-$2400~V across the two MCP plates, with a position resolution of $<0.1$~mm, temporal resolution of $<0.2$~ns, and a rate capability of $1$~MHz~\cite{PS_MCP_manual}.

\begin{table}[ht]
\small
\captionsetup{width = 14cm}
\caption[Typical DC biases applied to CPT electrodes]{\label{tab:cpt_dc_biases}Typical DC biases applied to the CPT electrodes}
\centering
\begin{tabular}{@{\extracolsep{4pt}}c|ccc}
\hline 

\multirow{2}{*}{Electrodes} & \multicolumn{3}{c}{Voltages [V]} \\ 
\cline{2-4}
 & Capture & Trap & Eject \\ 

\hline
    
Bottom correction tube & $-$17.5 & \phantom{$-$}0.9413 & \phantom{$-$}0.9413 \\
Bottom end cap & $-$13.5 & $-$3.0836 & $-$3.0836 \\

Bottom correction ring & $-$8.9622 & $-$8.9622 & $-$8.9622 \\
Ring &  $-$12.7206 & $-$12.7206 & $-$12.7206  \\

Top correction ring & $-$8.9622 & $-$8.9622 & $-$8.9622 \\
Top end cap & $-$3.0836 & $-$3.0836 & $-$13.5 \\
Top correction tube & \phantom{$-$}0.9413 & \phantom{$-$}0.9413 & $-$17.5 \\

\hline 
 \end{tabular}
\end{table}

\section{Mass measurement : Phase-imaging ion-cyclotron resonance (PI-ICR)}\label{sec:mass_meas}

\subsection{PI-ICR at the CPT}\label{subsec:piicr cpt}

\begin{figure}
\centering
\subfloat[]{\includegraphics[width=0.85\textwidth]{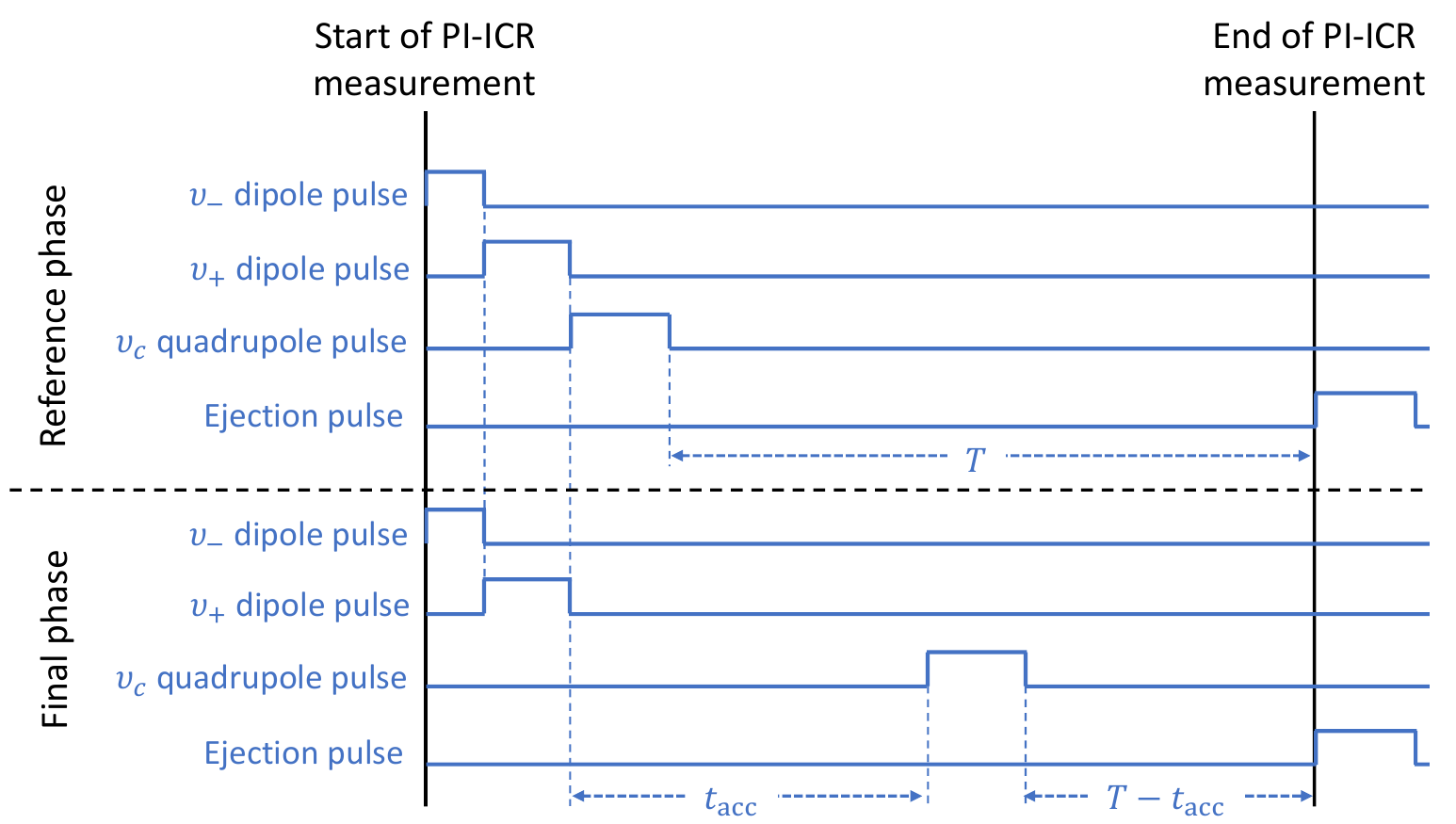}\label{fig:piicr_pulses}}\\
\vspace{2mm}
\subfloat[]{\includegraphics[width=0.45\textwidth]{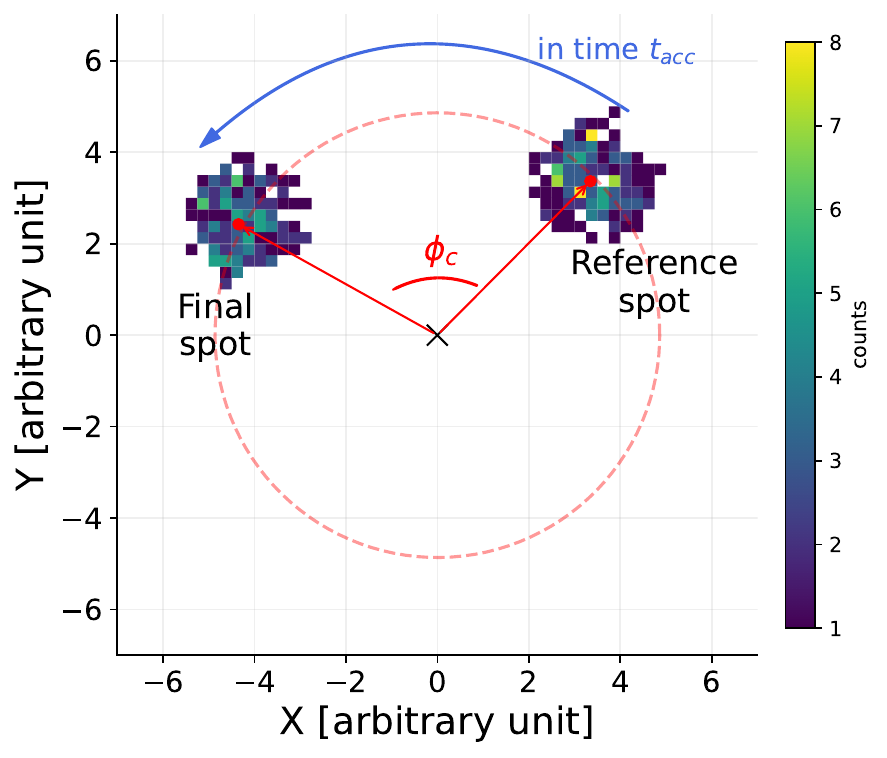}\label{fig:piicr_XY}}
  \subfloat[]{\includegraphics[width=0.52\textwidth]{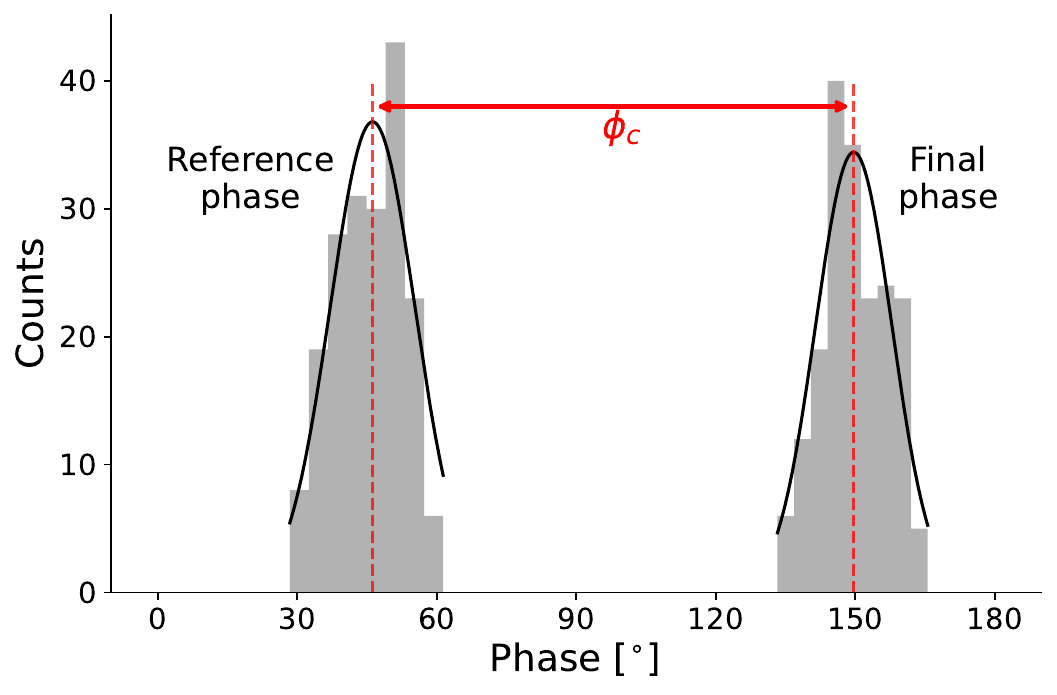}\label{fig:piicr_phase}}
\captionsetup{width = 12.5cm}
\caption[PI-ICR Illustration]{\label{fig:piicr_explanation}Illustration of the PI-ICR technique used at the CPT. Panel (a) shows the RF excitation schemes used during a PI-ICR measurement of reference (top) and final (bottom) phases. 
In panel (b), simulated reference and final spots are shown superimposed on one spectrum. The ions acquire an excess phase shown by the angle $\phi_{c}$ between the reference to the final spot in time $t_{\text{acc}}$. The \textit{spot-centroids}, obtained from clustering and fitting \cite{thesis_cpt_ray_phd_long, weber2021clustering}, is shown by the red dots, while the trap center is denoted by a black cross. The red dotted circle represents a hypothetical revolution-path the ions take post their excitement by the $\nu_{+}$ dipole pulse. In panel (c), the corresponding phase projections are shown along with fits to the data.}
\end{figure}

PI-ICR relies on determining a nuclide's $\nu_{c}$ via a ``sideband frequency'' measurement~\cite{eliseev_piicr_APB_long_2014}, where $\nu_{c} = \nu_{+} + \nu_{-}$ for an ideal Penning trap, $\nu_{-}$ and $\nu_{+}$ being the ion's magnetron and reduced cyclotron frequency respectively. This is done by obtaining the phases corresponding to the two radial motions acquired over periods of no excitation. 
The details of PI-ICR implementation at the CPT are described in Ref.~\cite{orford_nimb_piicr_cpt_2020}. 
One $\nu_{c}$ measurement cycle consists of two phase measurements: a reference phase ($\phi^{\text{ref}}$) and a final phase ($\phi^{\text{final}}$). Each of these phase measurements involves application of three RF excitations for each trap cycle: a $\nu_{-}$ dipole centering pulse, a $\nu_{+}$ dipole pulse to increase the radius of the corresponding motion and excite the ions to an orbit, and finally a $\nu_{c}$ quadrupole pulse to convert the ions’ fast $\nu_{+}$ motion to the slow $\nu_{-}$ motion for detection before they are ejected from the trap. 
This is repeated for multiple trap cycles to obtain enough ion-hits forming spots corresponding to the measured reference or final phase. The total trap time, durations of all three excitations, and timing structure of the two dipole excitations are kept the same for both the reference and the final phase measurements. The time of application of the $\nu_{c}$ pulse differentiates $\phi^{\text{ref}}$ and $\phi^{\text{final}}$. For the reference phase, the $\nu_{c}$ pulse is applied right after the end of the $\nu_{+}$ pulse, so the ions acquire $\nu_{-}$ phase in the time ($T$) between the conversion and the ejection pulse. 
As a result, the reference phase, in principle, is a $\nu_{-}$ phase. 
For the final phase, there is a delay ($t_{\text{acc}}$) between the two pulses, and the ions accumulate a combination of strongly mass-dependent $\nu_{+}$ phase over time $t_{\text{acc}}$, and $\nu_{-}$ phase over time $T-t_{\text{acc}}$. This means all beam species get resolved over a large-enough $t_{\text{acc}}$ for the final spots. 
The pulse schemes used for reference phase and final phase measurements are shown in Fig.~\ref{fig:piicr_pulses}, while a simulated spectrum of a reference and a final spot is shown in Fig.~\ref{fig:piicr_XY}.
It should be noted here that the conversion pulse preserves the accumulated phases of the eigenmotions, while flipping their signs~\cite{eliseev_piicr_APB_long_2014, kretzschmar_quad_exc_1}.
At the CPT, the ions' eigenmotions take place in a clockwise direction, which materialize in a counter-clockwise phase accumulation from the reference to the final spot, as shown in Fig.~\ref{fig:piicr_XY}.
Each reference and final spot is clustered and fitted using techniques described in Ref.~\cite{thesis_cpt_ray_phd_long, weber2021clustering} to obtain the corresponding \textit{spot-centroids} and hence the acquired phases. 
A phase projection of the two spots is shown in Fig.~\ref{fig:piicr_phase}.
The phase difference $\phi_{\text{total}} = \phi^{\text{final}} - \phi^{\text{ref}}$ can then be determined, and from it, one can obtain $\nu_{c}$, without considering any systematic effects, as:
\begin{equation}\label{eq:wc}
    \nu_{c} = \frac{\phi_{\text{total}}}{2 \pi t_{\text{acc}}} = \frac{\phi_{c} + 2 \pi N}{2 \pi t_{\text{acc}}}\; ,   
\end{equation}
where $\phi_{c}$ is the phase excess after completion of $N$ complete cyclotron revolutions in time $t_{\text{acc}}$ as shown in Fig.~\ref{fig:piicr_XY}.  

The determination of $N$ is crucial for conducting accurate measurements.
From Eq.~\ref{eq:wc}, a difference of $\Delta N$ turn would lead to a shift of $\Delta N/t_{\text{acc}}$~Hz in the determined $\nu_{c}$.  
For example, a difference of 1 turn for a measurement at $t_{\text{acc}} \sim 1000$~ms would lead to a relative shift of about 1.5~ppm in the determined $\nu_{c}$ for $^{133}$Cs$^{+}$.
Therefore, conducting accurate measurements, especially at large $t_{\text{acc}}$, require prior knowledge of a precise guess $\nu_{c}$ to unequivocally determine $N$.
For calibrant nuclides, whose masses are precisely known, $N$ is obtained from a guess frequency, determined from the reported mass in AME2020~\cite{ame2020_b_Wang2021}.
For target nuclides, a procedure is followed at the beginning of a measurement campaign to obtain the guess frequency.
A series of final phase measurements are conducted, starting with a small $t_{\text{acc}}$ ($\sim$5~ms) and then gradually increasing it, to determine more precise and accurate values for $\nu_{c}$. 
The guess $\nu_{c}$ is updated with every such measurement.
This is particularly necessary when measuring nuclides for which no prior experimental mass-data are available. 
For an impure beam, this method also allows for the identification of the $\nu_{c}$ frequencies of every species present in the beam, which is required for a phase correction (see Sec.~\ref{subsubsec:other sys effects}).
Once all beam species have been identified, a coarse $t_{\text{acc}}$ is selected ($\sim$100s of ms) such that all species in the beam are separated and clearly visible. 
Next a series of precise $t_{\text{acc}}$ values are selected, $\sim$100s of $\mu$s apart and spanning across roughly 1~-~1.5 $\nu_{-}$-periods, as required to correct for an initial $\nu_{-}$ motion (see Sec.~\ref{subsubsec:other sys effects}). 
Additionally, each $t_{\text{acc}}$ is chosen such that $|\phi_{c}|\leq10^{\circ}$, to account for the non-circular projection (see Sec.~\ref{subsubsec:elliptical projection}).
The time $T$ is chosen such that during a final phase measurement, $\nu_{+}$ phase accumulation over $t_{\text{acc}}$ and complete conversion to $\nu_{-}$ motion via the $\nu_{c}$ pulse can take place. 
Typically, $T$ is set 5~-~10~ms longer than $t_{\text{acc}}$, and remains unchanged for all reference and final phase measurements for a given $\nu_{c}$ measurement.

Once the $\nu_{c}$ of the target nuclide has been measured, $\nu_{c}$ of a calibrant species with well-known mass is measured using the same parameters to calibrate the magnetic field. 
The $\nu_{c}$ frequencies of both target and calibrant are measured following procedures described in Sec.~\ref{subsec:systematics} to minimize systematic effects, and a trap-center measurement is conducted as part of pre-measurement-campaign preparations. 
From the two $\nu_{c}$ measurements, the mass of the target ion is determined using Eq.~\ref{eq:mass}.

\subsection{Improving accuracy of PI-ICR measurements}\label{subsec:tuning}

The main factors that affect accuracy in PI-ICR measurements at the CPT include ensuring the target and calibrant ions probe the same magnetic field, a harmonic electric potential and resolution of PI-ICR spots. 
Effects relating to magnetic field inhomogeneities are minimized by fine-tuning the injection pulse and conducting an evaporative cleaning routine. With this procedure, the ions upon injection always get confined at the center of the trap in the axial direction where the magnetic field is largely homogeneous ($\Delta B \sim 0.1 \; \mu$T$/$mm). 

\begin{figure}
\centering
\includegraphics[clip,width=0.85\textwidth]{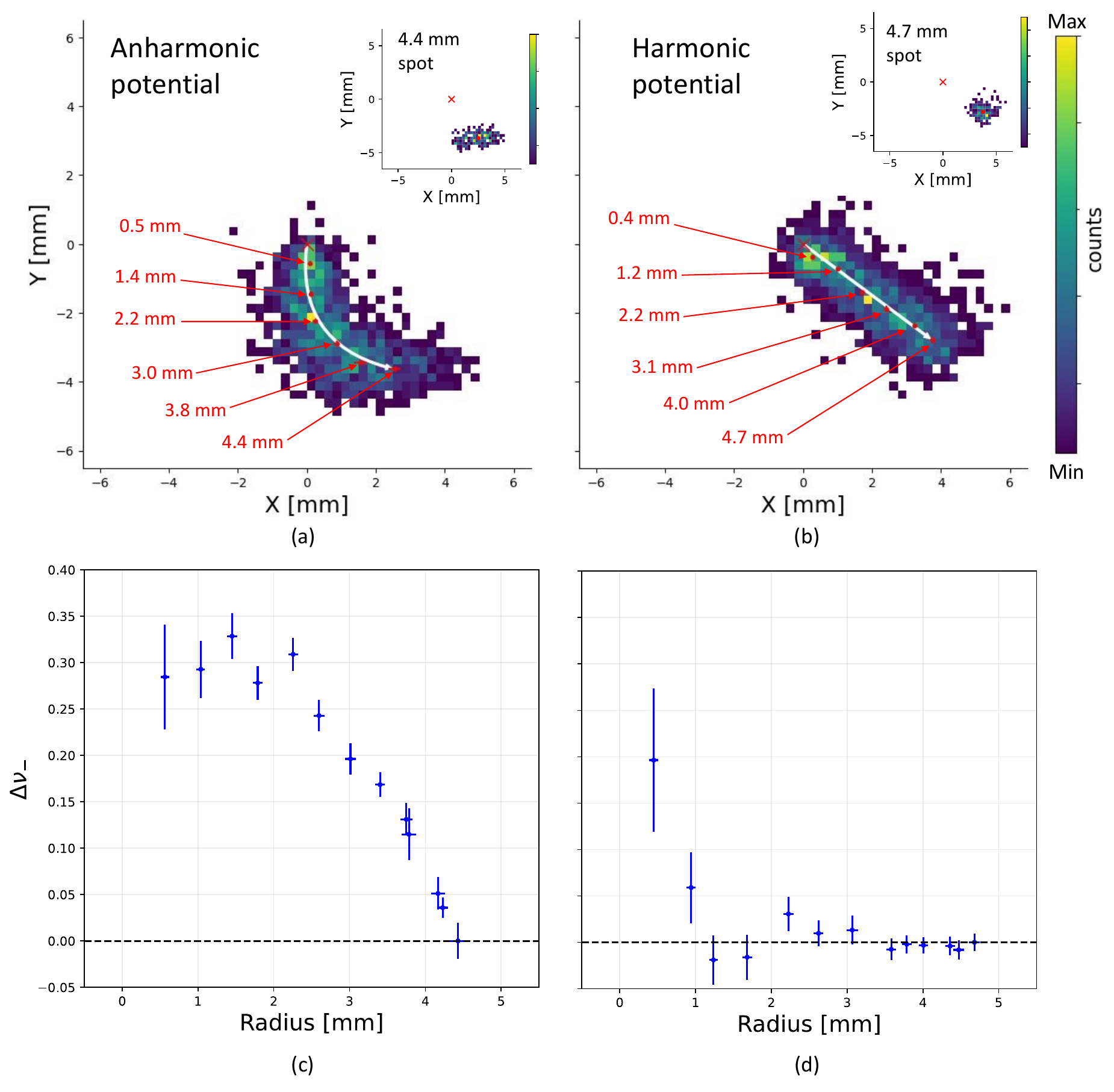}
\captionsetup{width = 12.5cm}
\caption[Tuning the bias applied to the correction tube]{\label{fig:tuning_CT}Creating a harmonic trap potential by tuning the correction tube potential. Left: an untuned correction tube potential of 1.5406~V leading to an anharmonic electric potential causing a misalignment in the $\nu_{-}$ spots at different radii and smearing of the spots (outermost spot shown on the inset) (in panel (a)), and the corresponding shifts in $\nu_{-}$ ( in panel (c)). Right: A tuned correction tube potential of 0.9413~V making the electric field harmonic and leading to aligned spots, with the outermost spot shown on the inset (panel (b)), and corresponding $\Delta \nu_{-}$ (panel (d)).}
\end{figure}

While for an ideal trap, $\nu_{c}$ is independent of the electric field, the practical aspects of the trap, including presence of apertures and segmented ring electrode, mean the harmonicity of the electric field plays an important role in precision measurement of $\nu_{c}$, as the eigenfrequencies $\nu_{+}$ and $\nu_{-}$ depend on the electric potential inside the trap. 
The shift in $\nu_{\pm}$ from the anharmonicities in the electric field, from non-zero octupolar ($C_{4}$) and dodecapolar ($C_{6}$) contributions alone, can be expressed as~\cite{brodeur2012titan_toficr}:
\begin{equation}\label{eq:anharmonic w- brodeur}
\begin{aligned}
\Delta \nu_{\pm} \approx \pm \frac{3\nu_{-}}{4d_{0}^{2}} \Bigg[C_{4}(r_{\pm}^{2}+2r_{\mp}^{2} - 2r_{z}^{2})
 + \frac{5C_{6}}{4d_{0}^2} \Big(-3r_{z}^{4} &+ 6 r_{z}^{2}(r_{\pm}^{2}+2r_{\mp}^{2}) \\
& - (r_{\pm}^{4}+3r_{\mp}^{4} + 6r_{+}^{2} r_{-}^{2})\Big) \Bigg] \;,
\end{aligned}
\end{equation}
where 
$d_{0}$ is the characteristic trap dimension given as $d_{0} = \sqrt{z_{0}^{2}/2 + r_{0}^{2}/4}$, and $r_{+}$, $r_{-}$, and $r_{z}$ are the radius of the $\nu_{+}$, $\nu_{-}$, and axial motions.
The correction tube electrodes can be used to minimize the $C_{4}$ and $C_{6}$ imperfections in the trapping potential~\cite{bollen1990accuracy, brodeur2012titan_toficr} and make the potential harmonic such that $\nu_{+}$ and $\nu_{-}$ are nearly independent of the amplitudes or radii of the motions. 
Using PI-ICR, this harmonicity can be optimized by conducting a series of reference phase measurements, following the pulse scheme shown in the top panel of Fig.~\ref{fig:piicr_pulses}.
The peak-to-peak voltage ($V_{\text{pp}}$) of the $\nu_{+}$ dipole excitation is changed for every measurement, while every other parameter including the duration of the excitations, the $V_{\text{pp}}$ of the $\nu_{-}$ and $\nu_{c}$ excitations, and the time $T$ is kept unchanged. 
Such a series of measurement results in the ions accumulating $\nu_{-}$ phases over time $T$ at different radii. 
With an optimized correction tube, the trapping potential will be harmonic resulting in a radial alignment of the reference spots.
On the other hand, a non-harmonic electric potential would lead to a radial dependence of $\nu_{-}$ and distortion of the PI-ICR spots~\cite{eliseev_piicr_APB_long_2014}. 
To optimize the correction tube voltage, the above procedure is repeated for a number of correction tube biases, while keeping the biases to the ring, two correction ring and two end-cap electrodes unchanged. 
The results are shown in Fig.~\ref{fig:tuning_CT} for measurements conducted at $T=250$~ms for an anharmonic potential, with the correction tube at 1.5406~V (panels (a) and (c)), and a harmonic potential, with the correction tube at 0.9413~V (Panels (b) and (d)). 
The top panels, (a) and (b), show a few selected reference spots at different radii with the outer most spots, typically used for mass measurement, shown on the inset. The corresponding shifts in the measured $\nu_{-}$ with respect to the outermost spot are shown in the bottom panels (c) and (d). 
With this optimization of the correction tube potential, the upper limit of the $C_{4}$ and $C_{6}$ can be estimated to $-4(10) \times 10^{-5}$, and $-2(3) \times 10^{-4}$ respectively.
As can be seen from the inset of Fig.~\ref{fig:tuning_CT} (a) and (b), a harmonic potential well will also lead to an undistorted image on the detector.

\begin{figure}
\centering
\includegraphics[clip,width=0.98\textwidth]{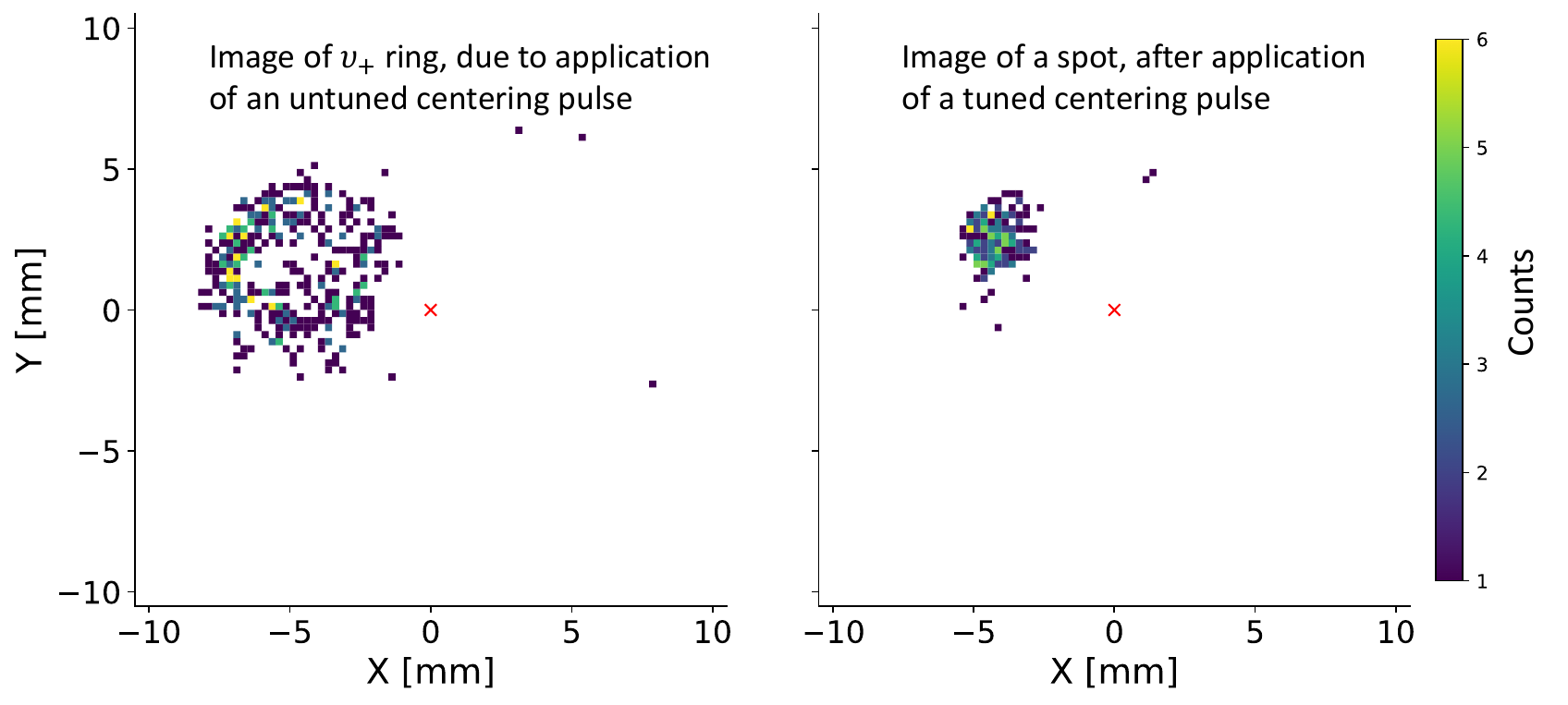}
\captionsetup{width = 12.5cm}
\caption[Tuning the centering pulse]{\label{fig:tuning_centering_pulse}Improvement in spot resolution by tuning the centering pulse. Left: An initial $\nu_{-}$ motion from an untuned centering pulse gets converted to a $\nu_{+}$ motion before ejection and gets imaged as a ring. Right: A tuned centering pulse minimizing the initial $\nu_{-}$ motion leading to observation of a spot-image at an orbit. The trap center is marked by the red cross.}
\end{figure}

The $\nu_{-}$ dipole `centering' pulse is applied so that the ions are centered before they are excited to a radius by the $\nu_{+}$ dipole pulse. An untuned centering excitation means the ions would possess a significant initial $\nu_{-}$ motion which continues after the $\nu_{+}$ excitation and gets converted to the fast $\nu_{+}$ motion by the $\nu_{c}$ pulse prior to ejection. As a result, the observed image becomes that of a ring instead of a spot. This is shown in Fig.~\ref{fig:tuning_centering_pulse}, where a four-parameter tuning of the amplitude, phase, duration and time-of-application of the $\nu_{-}$ pulse ensured detecting the image of a spot (right) instead of a ring (left).  
It should be mentioned here that a well-tuned $\nu_{-}$ pulse at the CPT means a negligible but non-zero initial $\nu_{-}$ motion which is adequate for spot optimization purposes, but is not sufficient to eliminate a sinusoidal dependence of the measured $\nu_{c}$ on $t_{\text{acc}}$, as described in Ref.~\cite{orford_nimb_piicr_cpt_2020} and briefly outlined in Sec.~\ref{subsubsec:other sys effects}.

\subsection{Systematic Effects}\label{subsec:systematics}

A number of systematic effects have been observed during PI-ICR implementation at the CPT. Some of these have been outlined in Ref.~\cite{orford_nimb_piicr_cpt_2020}, and continued to be refined. A few other effects like dependence on the angle of measurement of the target and calibrant species, characterization of the magnetic field and ion-ion interaction have been observed and studied since Ref.~\cite{orford_nimb_piicr_cpt_2020}.
For completeness,  a brief overview of all the effects is provided here.

\subsubsection{Non-circular projection}\label{subsubsec:elliptical projection}

\begin{figure}
\centering
\includegraphics[clip,width=0.98\textwidth]{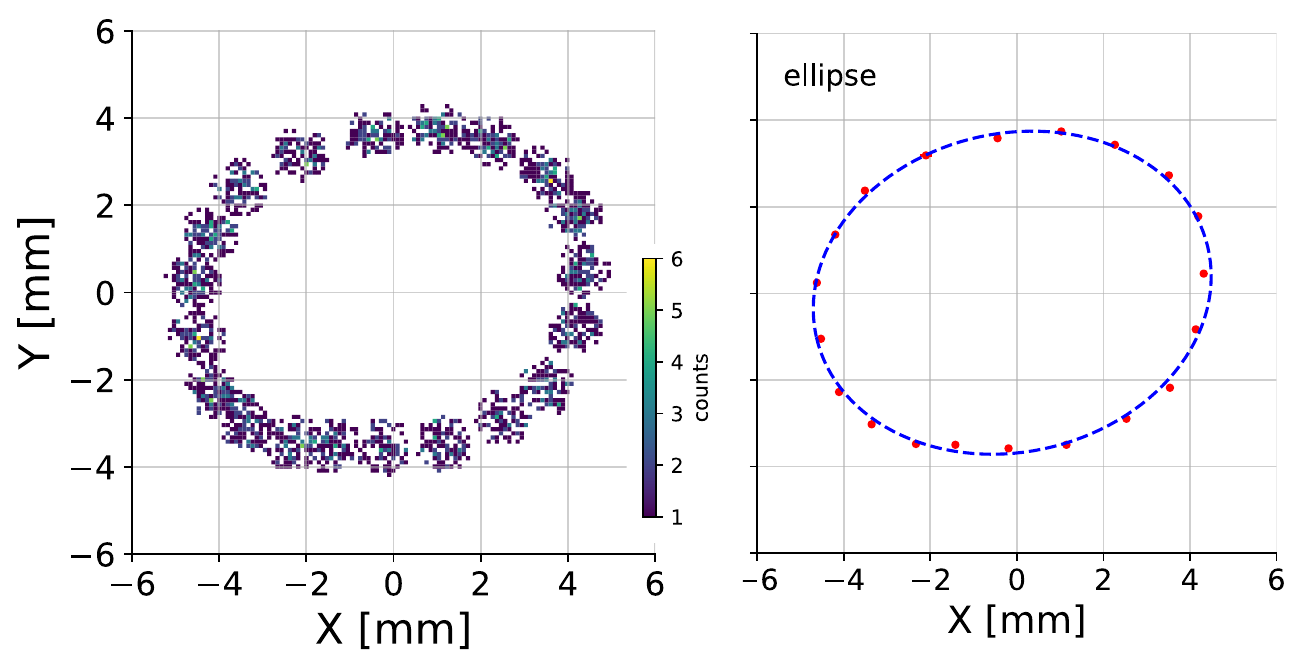}
\captionsetup{width = 12.5cm}
\caption[Projection of the $\nu_{c}$ phase from the trap to the detector]{\label{fig:phase projection}Projection of the $\nu_{c}$ phase from the trap to the detector. (Left) Each spot represents a $\nu_{c}$ phase acquired by $^{133}$Cs$^{+}$ ions over 6 $\mu$s. (Right) The spots are clustered and the spot centers (red dots) are fitted to an elliptical model with an ellipticity of 0.61(2).}
\end{figure}

Effects arising from a non-circular projection from the trap onto the detector have been briefly discussed in Ref.~\cite{orford_nimb_piicr_cpt_2020}. Figure~\ref{fig:phase projection} shows a $\nu_{c}$ phase walk conducted with $^{133}$Cs$^{+}$ ions with a series of $t_{\text{acc}}$ around 234~ms. Each spot represents a phase acquisition incremented by 6~$\mu$s. As can be seen, an elliptical fit best represents the data. 
Since PI-ICR relies on measuring the angular difference between two spots, the location of the spots around this ellipse would have a significant effect on the accuracy of the measurement.
The elliptical mapping could be due to a misalignment between the magnetic field axis and the trap axis. The effect can be reproduced in SIMION by tilting the magnetic field by $<$1$^{\circ}$~\cite{thesis_cpt_orford_phd_long}.  There were two ways to solve this issue: re-shimming the magnet such that the field axis coincides with the axis of the trap, or bringing the trap out, and machining and realigning its axis with the field axis.  
Re-shimming the magnet would need to be accompanied with a recalibration of the whole system, meaning a significant time would have been spent on preparing the system, with no guaranty of success in terms of the correct alignment.
The second method would have needed precision machining and again a significant period without operation.
Since the spontaneously fissioning Cf source in CARIBU is always weakening, both of these were postponed.
The problem, however, was minimized to within the typical  statistical limit of around 2~ppb by limiting the angular difference between the reference and final phases to $\leq 10^{\circ}$~\cite{thesis_cpt_orford_phd_long}. To achieve precision of $<$1~ppb, significant preparation is undertaken before a measurement campaign to limit $|\phi_{c}|$ to within $1^{\circ}$. Another effect relating to this elliptical projection has been recently investigated, where a difference between the angle of measurement of the target and that of the calibrant species would lead to a shift in the measured mass~\cite{liu2024precisemassmeasurement108}. 
To account for this angular dependence, the measured $\nu_{c}$ ratio $r$ is corrected using an empirical equation: $r^{'} = r \; \Theta(\theta_{1},\theta_{2})$, where 
$\Theta(\theta_{1},\theta_{2}) = (1+b \cos(\theta_{1} + \theta_{0}))/ (1+b \cos(\theta_{2} + \theta_{0} )$ with $\theta_{1}$ and $\theta_{2}$ being the angles of measurement of the target and the calibrant species. The constants $b$ and $\theta_{0}$ were obtained from offline measurements conducted with ions from SIS as
$b = 3.89(32) \times 10^{-8}$ and $\theta_{0} = 0. 80(6)$~rad~\cite{liu2024precisemassmeasurement108}.

\subsubsection{Electric field drift}\label{subsubsec:electric field}

\begin{figure}
\centering
\includegraphics[clip,width=0.99\textwidth]{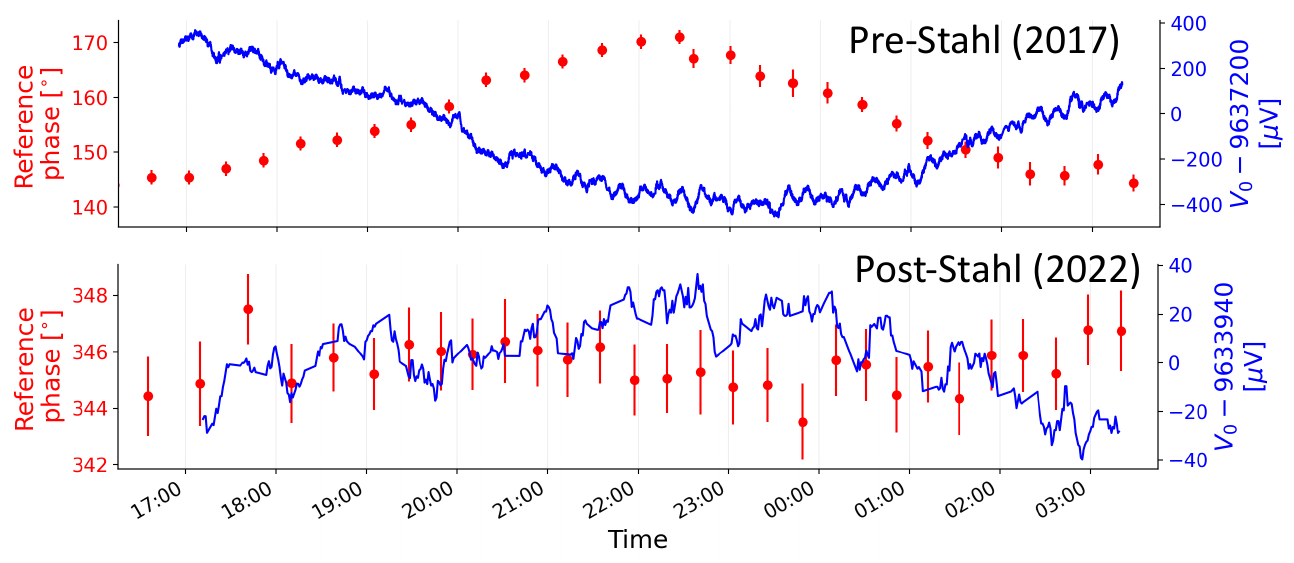}
\captionsetup{width = 12.5cm}
\caption[Electric field stability of the CPT]{\label{fig:electric_field_stability}Improvement in stability of the trap depth ($V_{0}$) and corresponding $\phi^{\text{ref}}$-drifts over 10 hours with the installation of the Stahl voltage supply}
\end{figure}

It can be shown that $\nu_{-}$ is proportional to $V_{0} / B$, where $V_{0}$ is the depth in the trapping potential, given by the potential difference between the end cap and the ring electrodes. Since, $\phi^{\text{ref}}$ is primarily a $\nu_{-}$ phase that the ions acquire over time $T$, the stability of the trap power supply is of paramount importance for precision measurements using PI-ICR. This effect was briefly discussed in Ref.~\cite{orford_nimb_piicr_cpt_2020}, and could lead to mass shifts of up to $\sim$0.1~ppm in extreme cases. The ``new'' power supply mentioned in Ref.~\cite{orford_nimb_piicr_cpt_2020}, an ultra-high precision voltage supply from Stahl Electronics (4 channel, $\pm$14~V BS Series)~\cite{stahl_cpt_ps} with extremely low temperature dependency ($10^{-6} \Delta$V$/$V per Kelvin) and temporal voltage fluctuations ($< 10^{-7} \Delta$V$/$V per minute), has since been thoroughly tested and is currently used as a dedicated supply to bias all the trap electrodes. The improvement in stability from this Stahl power supply is shown in Fig.~\ref{fig:electric_field_stability}. Before the installation of this new power supply, a $\sim$25$^{\circ}$ shift in $\phi^{\text{ref}}$ could be acquired over a $\sim$685~ms duration, and now, it has been reduced to $\sim$3$^{\circ}$ over a similar timeframe. As an additional precaution, reference phase data is acquired every few minutes or once per final file, then either the reference closest in time to a final file is considered for that final file, or interpolated \textit{true} references are determined and used. This has allowed for trapping the ions over larger times ($\sim$1~s) while minimizing the effects of electric field instabilities estimated to be within 2~ppb.

\subsubsection{Stability of the magnetic field}\label{subsubsec:magnetic field}

\begin{figure}
\centering
\includegraphics[clip,width=0.99\textwidth]{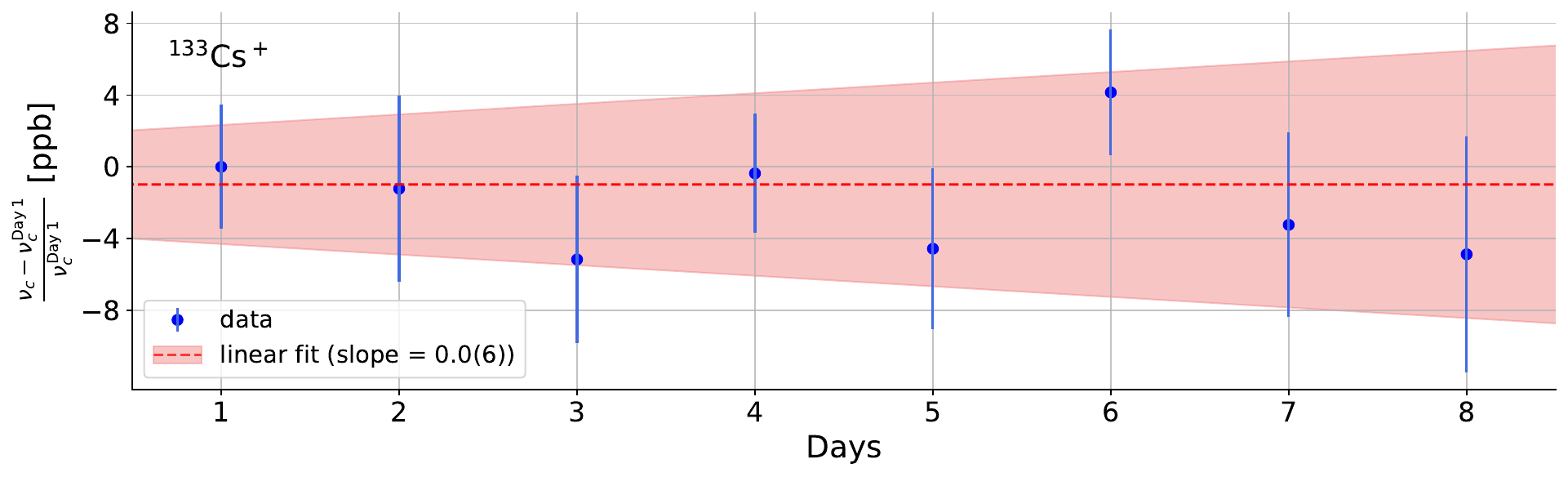}
\captionsetup{width = 12.5cm}
\caption[B field stability]{\label{fig:B_field_stability}Temporal drift of the magnetic field over 8 days shown as a relative difference in the measured cyclotron frequencies with respect to that measured on day 1 in ppb.}
\end{figure}

As can be seen from Eq.~\ref{eq:penning}, stability of the trap's magnetic field is critical for accurate measurements. 
The coils of the solenoid have a self-shielding design to reduce potential external field fluctuations~\cite{guy_cpt_first_1997, guy_cpt_anl_first_2001}.
Spatial drifts are addressed by ensuring the ions probe the same magnetic field in all three dimensions. 
To minimize temporal drifts, the conditions that could affect the stability, like the temperature of materials in the bore and the He recovery pressure in the magnet cryostat, are controlled~\cite{guy_cpt_first_1997, guy_cpt_anl_first_2001}.
Additionally, during measurements, field-calibrations are done as close in time as possible to a target-nuclide measurement. 
Figure~\ref{fig:B_field_stability} shows results of $\nu_{c}$-measurements conducted to gauge the stability of the magnetic field using $^{133}$Cs$^{+}$ ions.  The study was done over a period of eight days, which is the typical duration of experimental campaigns at the CPT. The measured cyclotron frequencies were fitted to a linear model with a negative slope of 0.0(6)~ppb$/$day. 
This result is consistent with that of a similar test, conducted over a span of 8 months, about 9 years prior using the previously employed Time-Of-Flight Ion-Cyclotron-Resonance (TOF-ICR) technique~\cite{bollen1990accuracy,guy_cpt_anl_first_2001}, which found this drift to be $-$0.6~ppb/day~\cite{thesis_cpt_morgan_msc_long}.
During mass measurements, several calibrant measurements with ions from CARIBU and SIS are interleaved in between target-nuclide measurements. 
Typically, the calibrations are conducted at least once every day, or before and right after the conclusion of mass measurement of scarcely produced isotopes, which could take up to 2-3 days.
If the calibrant and the target species are not measured on the same day, possible mass shifts from magnetic field drifts are calculated and added in quadrature to the measured uncertainties.

\subsubsection{Ion-ion interaction}\label{subsubsec:ion ion interaction}

\begin{figure}
\centering
\includegraphics[clip,width=0.99\textwidth]{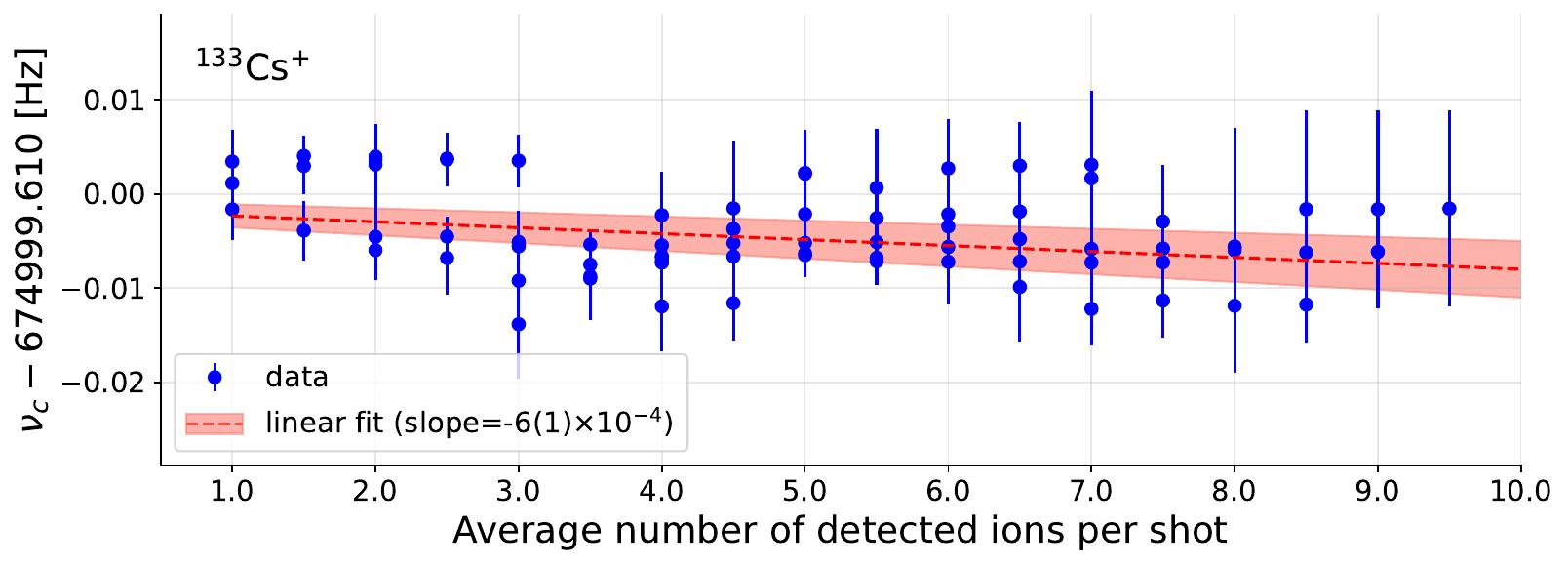}
\caption[count class analysis]{\label{fig:count_class_analysis}Drift in the measured $\nu_{c}$ from ion-ion interaction shown as average number of detected ions on the PS-MCP.}
\end{figure}

At the CPT, the number of trapped ions is controlled by limiting the rate of incoming ions into the trap.
The effect of ion-ion interaction was studied by conducting $\nu_{c}$ measurement of $^{133}$Cs$^{+}$ ions from SIS using five datasets with varying ion rates between 0.5~pps to 10~pps on the PS-MCP with a $\sim$70\% detector efficiency, keeping all other conditions identical.  
The data was analysed using multiple ranges of \textit{ion-cut}\footnote{\textit{ion-cut} considers only the data involving a user-defined range for number of ion-hits detected simultaneously on the PS-MCP.} condition.
This is shown in Fig.~\ref{fig:count_class_analysis}, along with a linear fit with a slope of $-6(1)\times10^{-4}$~Hz$/$ions detected.
During measurement campaigns, the ion rates are carefully maintained at $\sim$1-2~pps.
For most species from CARIBU, this was not a challenge, as they were detected at the CPT with rates $\leq 1$~pps.
For other species that are more abundantly produced including the stable ions from CARIBU (namely $^{84}$Kr$^{+}$, $^{86}$Kr$^{+}$ and C$_{6}$H$_{6}^{+}$) or ions from SIS, this would be done by detuning one or two steerers upstream from the linear trap.
Furthermore, during offline analysis, the ion-cut condition is typically set to $1-4$, and care is taken such that same number of detected ions per shot are considered for the target and the calibrant species, minimizing this effect to around 1~ppb in mass measurements.

\subsubsection{Other systematic effects}\label{subsubsec:other sys effects}

A mass dependent phase accumulation occurs during the $\nu_{+}$ and $\nu_{c}$ excitation. For final phase measurements, this is incorporated in the resolved spots for all the individual species. For reference phases, this does not result in multiple spots as, due to the short duration of the $\nu_{+}$ and $\nu_{c}$ excitations ($\sim$100s of $\mu$s), the net accumulated phase is smaller than the FWHM of the reference spot. This means the measured reference phase is a weighted average of the reference phases of all beam constituents on their relative populations.
To correct for this, a phase-correction 
($\phi_{\text{corr.}}$) given as~\cite{orford_nimb_piicr_cpt_2020}:
\begin{equation}\label{eq:phase correction}
\phi_{\text{corr.}} = 2 \pi t_{\text{exc}} \sum_{i} \chi_{i} (\nu_{c}^{i} - \nu_{c}) \; ,
\end{equation}
where $\chi_{i}$ and $\nu_{c}^{i}$ are the relative population fraction and cyclotron frequency of the $i$-th species, and $t_{\text{exc}}$ is the total duration of the $\nu_{+}$ and $\nu_{c}$ excitations, is applied. Due to the presence of the MR-TOF upstream, this typically results in an error of less than 2~ppb. 
The negligible but non-zero $\nu_{-}$ motion that the ions possess after the centering excitation as mentioned in Sec.~\ref{subsec:tuning}, is retained and gets coupled to the $\nu_{+}$ motion after the $\nu_{+}$ excitation is applied. 
This residual $\nu_{-}$ motion results in a sinusoidal dependence of measured phase $\phi_{c}$, and hence the frequency $\nu_{c}$, on $t_{\text{acc}}$~\cite{orford_nimb_piicr_cpt_2020}. To account for this effect,  a series of $\nu_{c}$ measurements is conducted by varying $t_{\text{acc}}$ over at least one $\nu_{-}$ period, and then fitted to a sinusoidal model 
~\cite{orford_nimb_piicr_cpt_2020}:
\begin{equation}\label{eq:sinefit}
\nu_{c}(t_{\text{acc}})=\frac{k_{1}}{t_{\text{acc}}} \sin \left(\nu_{-} t_{\text{acc}} + k_{2} \right) + \overline{\nu}_{c} \; ,
\end{equation}
where $k_{1}$ and $k_{2}$ are the amplitude and phase of the $\nu_{c}$ oscillation.
The baseline $\overline{\nu}_{c}$ obtained from the fit gives the true cyclotron frequency. For more scarcely produced isotopes, 1 - 2 $\nu_{c}$ measurements are conducted, about 1$/$2 $\nu_{-}$ period apart, and then the model parameters are characterized using an abundant isobar ion to deduce $ \overline{\nu}_{c}$ of the target species, while inflating the measured uncertainty.
Finally,  effects related to the difference in $A/q$ between the calibrant and measured nuclide were estimated to be $\leq$0.4(2)~ppb$/\Delta$u from measurements of isotopes of well-known masses from SIS and CARIBU, where $\Delta u = (A/q)-(A/q)^{\text{Cal}}$.
Detailed discussions of the mass dependent shift along with the angular dependence from Sec.~\ref{subsubsec:elliptical projection} are being prepared for a future publication.

\subsubsection{Summary}\label{subsubsec:summary}

A summary of systematic effects observed with PI-ICR at the CPT is shown in Table~\ref{tab:summary_systematics}.
All masses recently measured and reported were determined following these. The uncertainty from the \textit{sine-fit} model is incorporated in the statistical uncertainty of measurement, while that due to contaminant species is propagated from the individual components in the calculation. Furthermore, a systematic contribution of 3.0~ppb is added in quadrature to the statistical uncertainties to account for possible shifts due to electric field instabilities, ion-ion interactions, and non-circular projection from the trap to the detector, in addition to the magnetic field shift.

\begin{table}
\small

\centering
\captionsetup{width = 14cm}
\caption[Summary of systematic effects]{\label{tab:summary_systematics}Summary of systematic effects with their sources,  and methods implemented to minimize their effect.}
\begin{tabular} { p{5cm} | p{8cm} }
\hline


\makecell{Source} & \makecell{Methods to address (magnitude)}  \\

\hline

\makecell{Non-circular projection} &
$t_{\text{acc}}$ selected such that $|\phi_{c}| \leq 10^{\circ}$ ($< 2$~ppb).
Cyclotron frequency ratio corrected using the factor $\Theta(\theta_{1},\theta_{2})$, as discussed in Sec.~\ref{subsubsec:elliptical projection}. \\ 

\makecell{Electric field drift} &
Reference phases measured periodically and the one closest to each final file is used ($< 2$~ppb). \\ 

\makecell{Stability of the magnetic field} &
Calibration done several times during week-long campaigns, with the closest one, typically within a day, used for target-ion measurement (0.0(6)~ppb$/$day).\\

\makecell{Ion-ion interaction} &
Ion rates kept low and \textit{ion-cut} applied to ensure same number of detected ions per shot for both the calibrant and the target species (around 1~ppb). \\

\makecell{Phase correction for impure beam} &
All beam species and their relative population determined and a phase correction (see Eq.~\ref{eq:phase correction}) applied (typically $<$2~ppb).\\

\makecell{Residual $\nu_{-}$ motion} &
Series of final phase measurements conducted varying $t_{\text{acc}}$ over a $\nu_{-}$ period, data fitted to a sinusoidal model (see Eq.~\ref{eq:sinefit}), and true cyclotron frequency obtained from fit (standard error obtained from fit).\\

\makecell{Mass dependent shift} &
Ratio corrected as 0.4(2)~ppb$/\Delta$u. The shift is also added in quadrature to the uncertainty to be conservative. \\ [4pt]

\hline
\end{tabular}
\end{table}

\section{Outlook}\label{sec:outlook}

PI-ICR implementation at the CPT at CARIBU has been a fruitful endeavour measuring masses of more than 200 neutron-rich nuclides over six years, resulting in multiple publications and theses. PI-ICR has increased the measurement sensitivity of the CPT helping achieve ppb-level precision~\cite{thesis_cpt_ray_phd_long, hoff_mass_128mSb_PhysRevLett.131.262701, valverde2024precise, WSPorter_PhysRevC.110.034321} and record mass resolving power of $\sim$10$^{7}$~\cite{Mitchell_106Nb_PhysRevC.103.024323}, and allowing for measurement of some of the most weakly populated $^{252}$Cf-fission fragments~\cite{orford_prl_Nd-Sm_2018, thesis_cpt_ray_phd_long}. The CPT system is currently being transported to the upcoming N=126 Factory~\cite{SAVARD2020258_n126} in ATLAS at ANL.  While most of the system will be kept unchanged, a few upgrades, including a revamp of the control system and introduction of a Lorentz steerer~\cite{RINGLE_lorentz_200738} just before the trap for efficient on-center injection into the trap are planned. Although this marks the end of the CPT-at-CARIBU project, it points towards an exciting future in studying neutron-rich nuclides far from stability.

\section*{Acknowledgements}

This work was performed with the support of US Department of Energy, Office of Nuclear Physics under Contract No. DE-AC02-06CH11357 (ANL), the Natural Sciences and Engineering Research Council of Canada under Grant No. SAPPJ-2018-00028, and the US National Science Foundation under Grant No. PHY-2310059. This research used resources of ANL’s ATLAS facility, which is a DOE Office of Science User Facility.

\bibliographystyle{elsarticle-num2-3author.bst}

\bibliography{CPT_caribu_system_2024_V3}

\end{document}